\documentclass[a4paper]{jpconf}
\usepackage{graphicx}
\usepackage{epstopdf}
\usepackage{amsmath,amssymb}
\usepackage{axodraw}
\usepackage{color}

\def\beq{\begin{equation}}
\def\eeq{\end{equation}}
\def\beqa{\begin{eqnarray}}
\def\eeqa{\end{eqnarray}}

\begin{document}
\title{SM single-top production at hadron colliders}

\author{Pietro Falgari}

\address{Institute for Theoretical Physics and Spinoza Institute,\\ Utrecht University, 3508 TD Utrecht, The Netherlands}

\ead{p.falgari@uu.nl}

\begin{abstract}
We give an overview of the status of theoretical predictions for single-top production in the Standard Model.
We focus in particular on recent developments, including calculations of off-shell effects at next-to-leading 
order beyond the narrow-width approximation, all-order resummation of soft corrections and matching of 
next-to-leading order parton-level results to Monte Carlo parton showers.\\
\vspace{1 mm}
{\flushleft \emph{Preprint numbers:} ITP-UU-13/01,  SPIN-13/01.}
\end{abstract}

\section{Introduction}

The hadroproduction of a single top quark, first observed at the Tevatron \cite{Aaltonen:2009jj,Abazov:2009ii}, 
is a process of significant phenomenological relevance, providing 
informations complementary to those that can be obtained from top-quark pair production. Being mediated by 
electroweak-boson exchange, single-top production represents an unique window into the charged-current 
interactions of the top quark, enabling tests of the $V-A$ structure of the $Wtb$ vertex and a direct extraction
of the CKM matrix element $V_{tb}$, which is at the moment only indirectly constrained. Single-top production 
is also very sensitive to new-physics effects and anomalous couplings, whose strength can be assessed by a precise
measurement of the production cross sections. Furthermore, single-top production probes the bottom-quark 
parton distribution inside the proton, which at present is weakly constrained by other experimental data.         
For these reasons, the study of the production of a single top quark will be an 
important part of the physics programme of the Large Hadron Collider (LHC) at Cern, where several results
are already available \cite{Aad:2012ux, :2012dj, Chatrchyan:2012ep, :2012yva}. 

\begin{figure}[t!]
\begin{center}
\includegraphics[width=0.9 \linewidth]{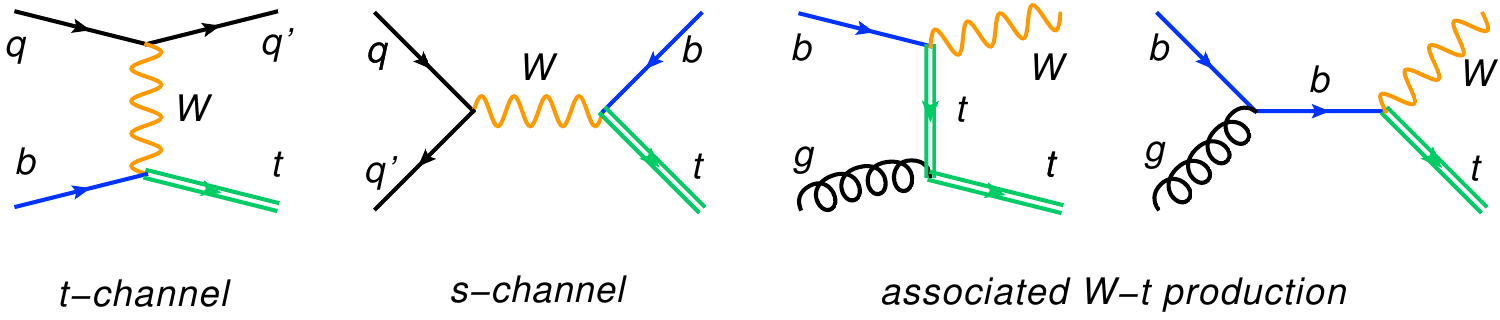}
\end{center}
\caption{Tree-level single-top production in the Standard Model.}
\label{fig:tree}
\end{figure}

In the Standard Model (SM) single-top production is customary divided into three production channels:
$t$-channel production, which involves the exchange of a space-like $W$ boson, $s$-channel production,
where the intermediate $W$ boson is time-like, and associated production of a top quark and a real $W$ boson.
The tree-level Feynman diagrams contributing to the three different channels are shown in Figure \ref{fig:tree}.  
Both at Tevatron and the LHC the $t$-channel process is the dominant one, accounting for about $63\%$ and 
$71\%$ of the cross section, respectively \cite{Bernreuther:2008ju} \footnote{The relative size of the three channels changes slightly
depending on the theoretical prediction used.}. At the Tevatron $s$-channel production is the second most important
channel ($\sim 30\%$) but it is negligible at the LHC, where $Wt$ production has the second largest cross section (about $25\%$). 
It has to be mentioned that the classification into three separate channels is delicate and somewhat
artificial, since, starting at next-to-leading order (NLO) in $\alpha_s$ $t$-channel and $s$-channel production mix with each
other. Furthermore, at NLO $Wt$ production interferes with $t\bar{t}$ production, which makes
the theoretical definition of two separate processes difficult and the extraction of the $Wt$ signal from the 
much larger $t\bar{t}$ background experimentally challenging. The prospects of a meaningful theoretical and experimental 
definition of the $Wt$ process were thoroughly investigated in \cite{Frixione:2008yi, White:2009yt}.     

NLO QCD predictions for single-top production in the approximation of a stable top have been available for 
more than ten years, both at the inclusive \cite{Bordes:1994ki, Stelzer:1997ns, Smith:1996ij, Giele:1995kr, Zhu:2002uj} 
and differential level \cite{Harris:2002md, Sullivan:2004ie}.
Electroweak and SUSY-QCD corrections have been also computed and found to be small, typically below $5\%$ of the tree-level result \cite{Beccaria:2008av, Macorini:2010bp}.  
More recently the relation between the five-flavour (5F) and four-flavour (4F) scheme at NLO has been investigated \cite{Campbell:2009ss,Campbell:2009gj}.
The two approaches differ in the treatment of the initial bottom quark in $t$-channel single-top production, which originates 
from a non-vanishing bottom PDF in the 5F scheme, and is generated by splitting of an initial gluon in the 4F scheme. 
In \cite{Campbell:2009ss,Campbell:2009gj} the two schemes were found to be in good agreement at NLO, except for distributions related to
the spectator $b$-jet, which in the $5$F scheme are effectively LO observables.  Beyond the stable-top approximation, NLO predictions for 
single-top production and decay have been computed in the Narrow Width Approximation (NWA) \cite{Campbell:2004ch, Cao:2004ky, Cao:2005pq, Campbell:2005bb, Heim:2009ku, Schwienhorst:2010je}, and were implemented in the numerical code MCFM \cite{Campbell:2012uf}. 
In this approach the top quark is produced on shell and let decay, with full spin correlations, to its final products.

Recent developments beyond a fixed-level NLO calculation for production and decay of an on-shell top
include the calculation of off-shell and non-factorizable corrections to $t$-channel and $s$-channel single top production,
soft-logarithm resummation
of the partonic cross sections
and the matching of NLO parton-level calculations to Monte Carlo parton showers. We will cover these topics in the following sections.  
New-physics effects in single-top production in extensions of the SM have also been widely investigated, but they are beyond the 
scope of this review.

\section{Off-shell and non-factorizable corrections to single-to production}

In the framework of the NWA NLO contributions are given by factorizable corrections to the
on-shell production and decay of the top quark, while non-factorizable contributions connecting initial- 
and final-state light partons, as well as off-shell and finite-width effects, are neglected. These terms are
small for the total cross section, of order of the top width-to-mass ratio $\Gamma_t/m_t \sim 1\%$, due to large cancellations between
virtual and real corrections. However non-factorizable corrections could be a priori large for exclusive observables, like arbitrary
kinematical distributions. This was investigated in \cite{Falgari:2010sf, Falgari:2011qa}, where off-shell 
and non-factorizable corrections to $t$-channel and $s$-channel production were computed.

The calculation of \cite{Falgari:2010sf, Falgari:2011qa} is based on an effective-field theory (EFT) description of the 
single-top production process \cite{Beneke:2004km}, built upon the hierarchy $\Gamma_t \ll m_t$. In this
approach contributions to the amplitude are divided into hard corrections, encoding physics at the 
large momentum scale $q \sim m_t$, and soft contributions, describing the long-distance physics 
associated with the low scale $q \sim \Gamma_t$. In the effective theory only soft modes are described by
dynamical fields, while hard contributions are encoded into the effective couplings (matching coefficients) of the EFT Lagrangian. 
In this framework an NLO calculation requires ${\cal O}(\alpha_s)$ corrections to the hard matching 
coefficients and one-loop soft contributions to the EFT matrix elements. These can be defined in terms of an 
expansion of full QCD diagrams in hard and soft momenta using the method of regions 
\cite{Beneke:1997zp}.  

\begin{figure}[t!]
\begin{center}
\includegraphics[width=0.34 \linewidth,angle=-90]{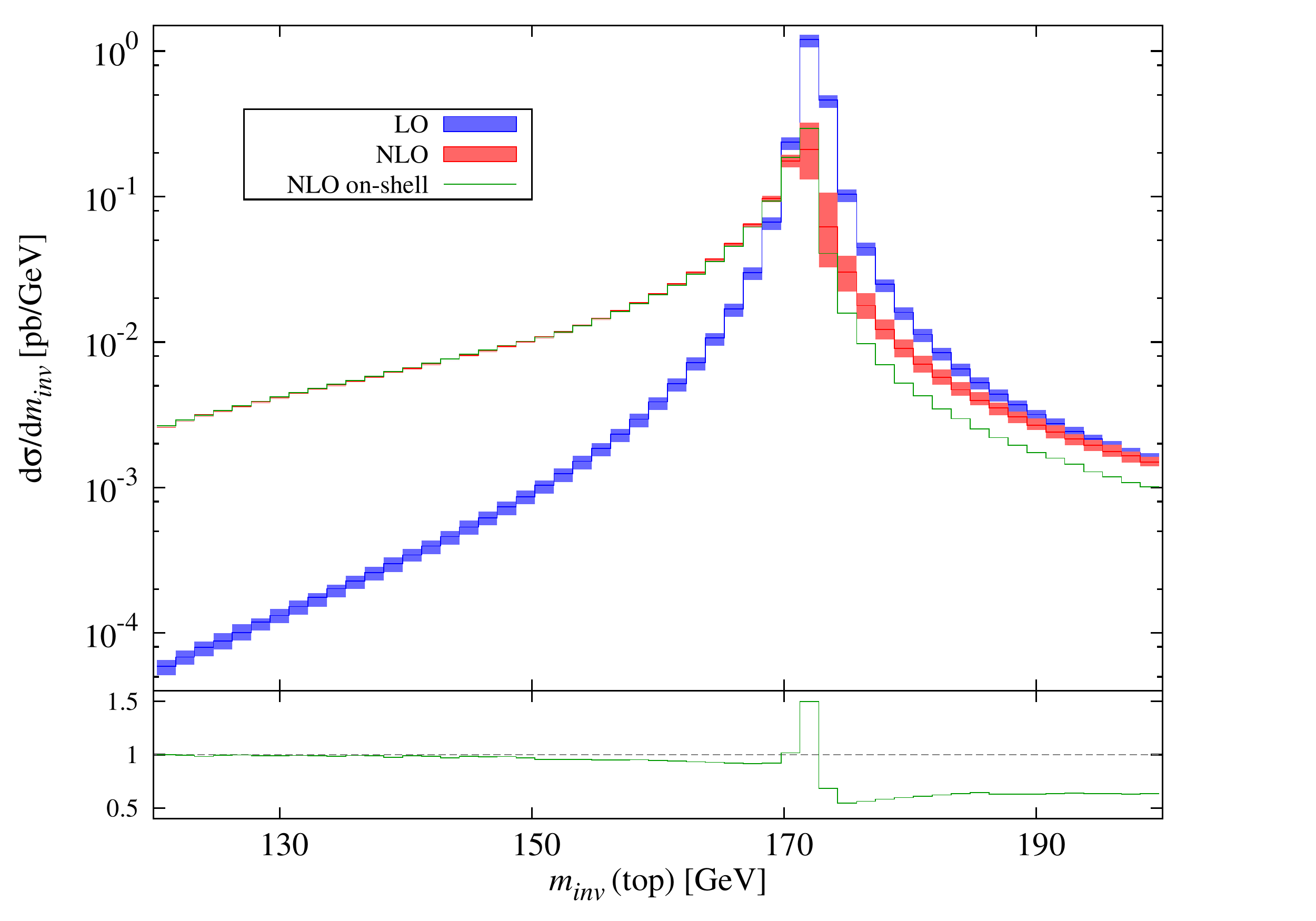}
\includegraphics[width=0.34 \linewidth,angle=-90]{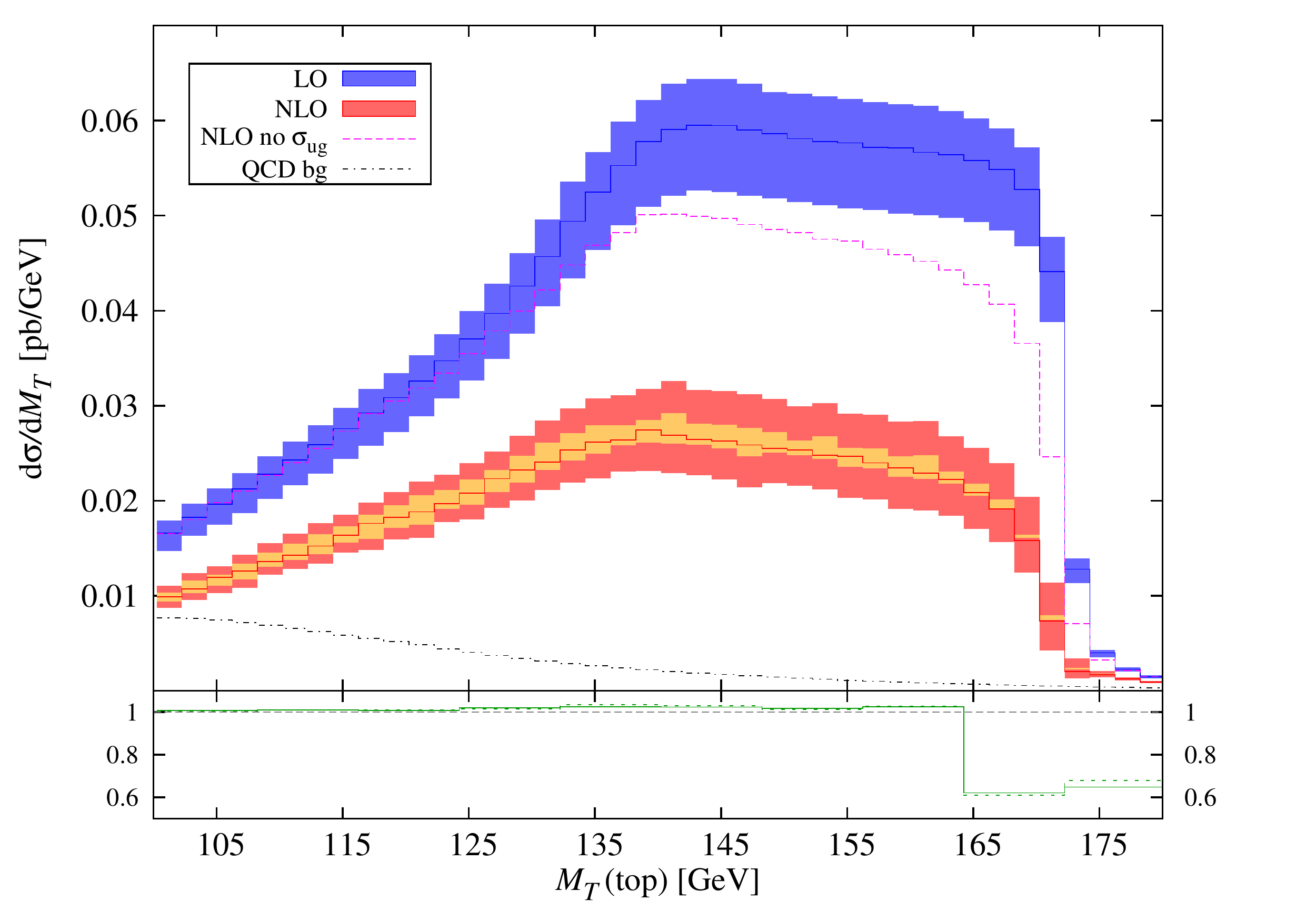}
\end{center}
\caption{Top-quark invariant-mass (left) and transverse-mass (right) distributions for $t$-channel single top production at the LHC with $\sqrt{s}=7\,$TeV
\cite{Falgari:2011qa}.
See the text for explanation.}
\label{fig:EFT_result}
\end{figure}
Two examples of the effect of off-shell and non-factorizable corrections are shown in Figure \ref{fig:EFT_result},
where the top invariant-mass and transverse-mass distributions for $t$-channel single-top production at a 7 TeV
LHC are given for $m_t=172\,$GeV \cite{Falgari:2011qa}. The blue and red curves represent the LO and NLO off-shell result, respectively, 
the bands the corresponding uncertainty obtained from scale variation. The green curve in the lower inset is the ratio of the
NLO on-shell prediction in the NWA to the off-shell result. One can seen that off-shell and non-factorizable effects are 
large around the peak of the invariant-mass distribution, with deviation of up to $50\%$ from the on-shell result. 
However these effects change sign around the peak, which explains the small corrections in the
transverse-mass distribution for $M_T<m_t$, where large cancellations occur due to averaging over different values of 
the invariant mass for a fixed $M_T$. These cancellations are less effective 
close the the distribution edge $M_T \sim m_t$, due to phase-space restrictions. In this region off-shell effects are large, 
about $40\%$ of the NLO on-shell result, and are necessary for a reliable theoretical prediction of the distribution.
 
\section{Resummation of soft logarithms}

Single-top production observables contain logarithmic contributions $\alpha_s^m [\ln^k s_4/s_4]_+$ (with $k=0,...,2 m-1$) 
related to suppression of soft-gluon emission near kinematical thresholds, with $s_4$ 
the (observable dependent) kinematical variable that vanishes at threshold. Such terms can give a sizeable contribution 
to the cross sections, and all-order resummation of these threshold terms can be used as a mean to 
improve fixed-order predictions. For single-top production, this has been studied separately by two different groups, using a formalism
based on resummation in Mellin-moment space \cite{Kidonakis:2006bu, Kidonakis:2007ej} and a framework involving 
soft-collinear effective theory (SCET) and renormalization group evolution equations \cite{Zhu:2010mr, Wang:2010ue}.

Both methods are based on the factorization of the cross section at threshold into a hard function $H$ describing the
short-distance scattering process, a soft function $S$ encoding soft radiation and additional terms accounting for 
collinear emission from initial- and final-state particles. The all-order resummation of soft contributions in $S$ is
controlled by matrix-valued soft anomalous dimensions $\Gamma_s$, which have been computed up to two loops \cite{Kidonakis:2010tc,Kidonakis:2010ux,Kidonakis:2011wy,Becher:2009kw}. 

Explicit results for the NNLL resummation of the single-top production cross section are available for the
three separate channels in the Mellin formalism \cite{Kidonakis:2012rm}, and for $t$-channel \cite{Wang:2010ue} 
and $s$-channel production \cite{Zhu:2010mr}
in the SCET approach. Both formalisms find small resummation effects, of order of few percents, in $t$-channel single-top 
production. On the other hand, in $s$-channel production the results of the two groups show bigger
discrepancies, with far larger soft effects ($13\%-15\%$) found in the Mellin approach \cite{Kidonakis:2010tc}. 
This discrepancy has not been investigated yet. However it should be noted that in the two approaches
different kinematical variables $s_4$ are chosen, so that sets of different logarithms are resummed. 
While the two calculations should give formally equivalent results at NNLL accuracy for the total cross section, 
power-suppressed contributions, which are not controlled by resummation, could be numerically 
important and be responsible for the observed differences.     

\section{Matching of NLO results to Monte Carlo parton showers}

While a fixed-order NLO calculation gives an accurate description of wide-angle 
radiation, in the low $p_T$ region multiple collinear emission becomes important 
and has to be taken into account. Furthermore, to realistically describe physical 
final states, a parton-level calculation has to be interfaced to 
some kind of hadronization model describing the evolution of energetic coloured partons
into colour-singlet hadronic states. This can be done in the framework of Monte Carlo parton 
showers (MCPS), which gives a probabilistic description of multiple collinear splitting and can be easily 
interfaced to one's preferred hadronization model.     
Recently a lot of effort has been put into the matching of NLO fixed-order calculations to Monte 
Carlo parton showers, in an attempt to obtain a framework in which both large-angle radiation
at NLO and all-order collinear emission are correctly described. This requires addressing 
double-counting issues in the collinear region, which has been consistently done in two different frameworks, 
MC@NLO \cite{Frixione:2002ik} and POWHEG \cite{Frixione:2007vw}.  The production of a single top have been
implemented in both approaches \cite{Frixione:2005vw, Alioli:2009je,Re:2010bp}.        
\begin{figure}[t!]
\begin{center}
 \includegraphics[width= 0.46 \linewidth]{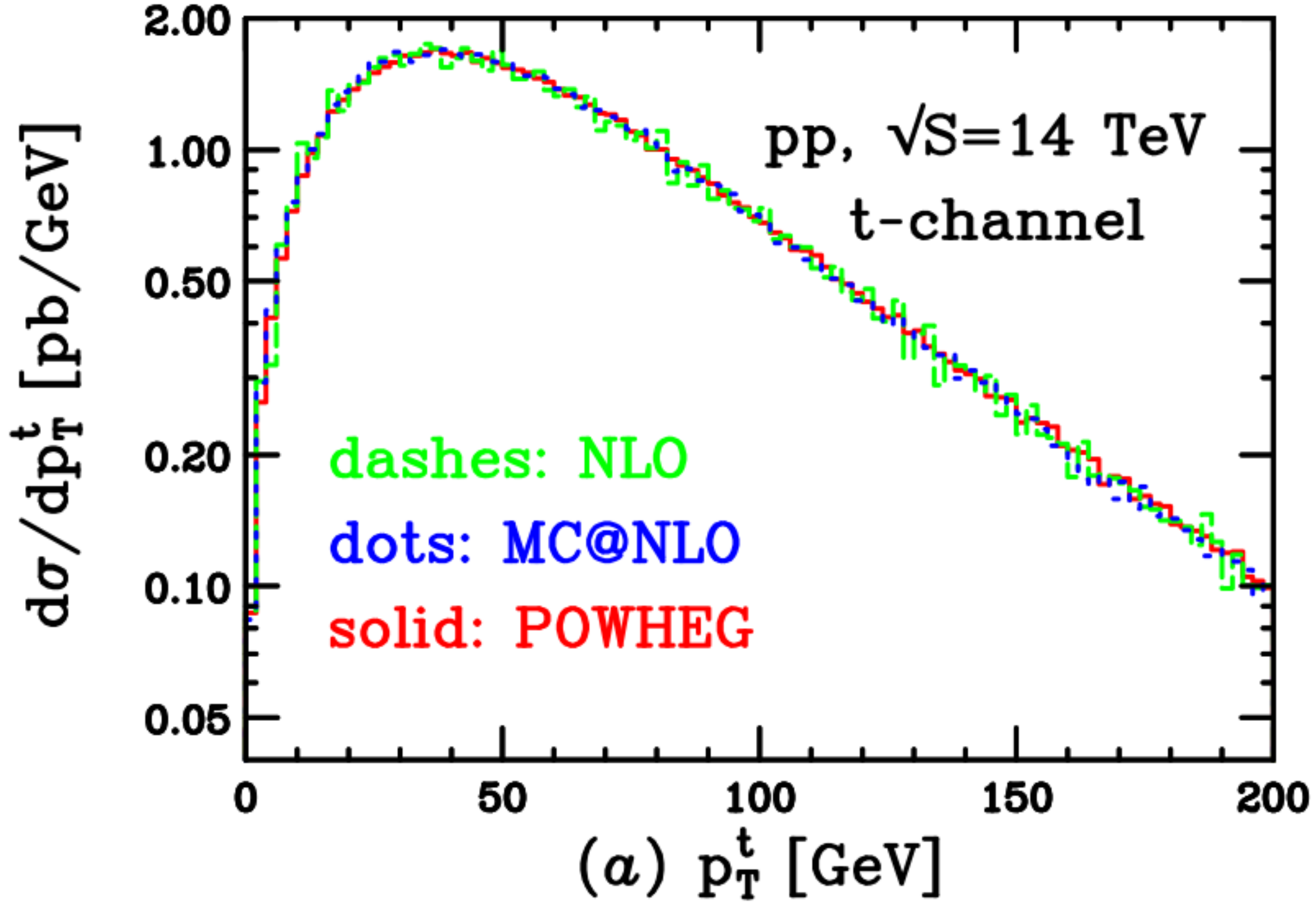}
 \includegraphics[width= 0.45 \linewidth]{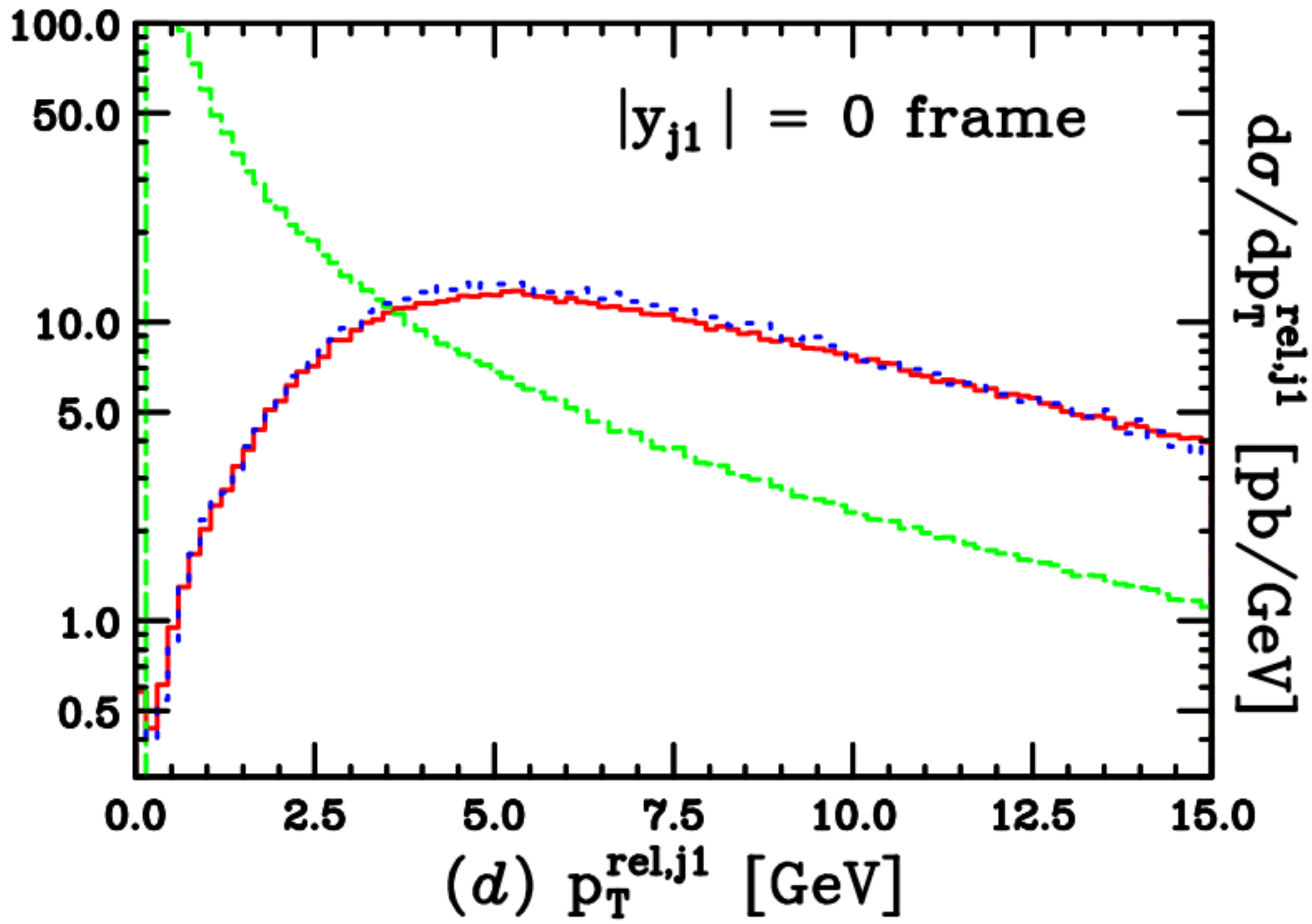}
\end{center}
\caption{Comparison of fixed-order NLO results for $t$-channel single-top production at the LHC with MCPS results obtained with POWHEG and MC@NLO (plots taken from \cite{Alioli:2009je}). See the text for explanation.}
\label{fig:NLO_vs_MCPS}
\end{figure}

Figure \ref{fig:NLO_vs_MCPS} shows the comparison of the two NLO Monte Carlo showers with a fixed-order
NLO result for the specific case of $t$-channel single-top production at the LHC with $\sqrt{s}=14\,$TeV. The two 
plots (taken from \cite{Alioli:2009je}) represent the top-quark transverse-momentum distribution (left) and the distribution
of the relative momentum of all the partons clustered inside the hardest jet, $p_T^{\text{rel},j_1}$, in the reference frame in which the 
rapidity of the jet is zero. In both cases the good agreement between the results of the two NLO parton showers is
evident. For the top transverse-momentum distribution one can also note the good agreement between the fixed-order
NLO prediction and the showered results. This is not the case for $p_T^{\text{rel},j_1}$, where the parton-shower and fixed-order
results show stark differences. This is expected, since the total relative momentum gives a measure of the spreading of the 
hardest jet, whose correct description requires the inclusion of multiple collinear emission when $p_T^{\text{rel},j_1} \rightarrow 0$.     
Comparisons for the $s$-channel process and associated $tW$ production \cite{Re:2010bp} have shown a similar good agreement between 
MC@NLO and POWHEG, and confirm the importance of a NLO parton-shower treatment to correctly describe observables in both the small
and large transverse-momentum regions.            

\begin{figure}[t!]  
\begin{center}
\begin{minipage}{0.5 \linewidth}
\includegraphics[width= 0.95 \linewidth]{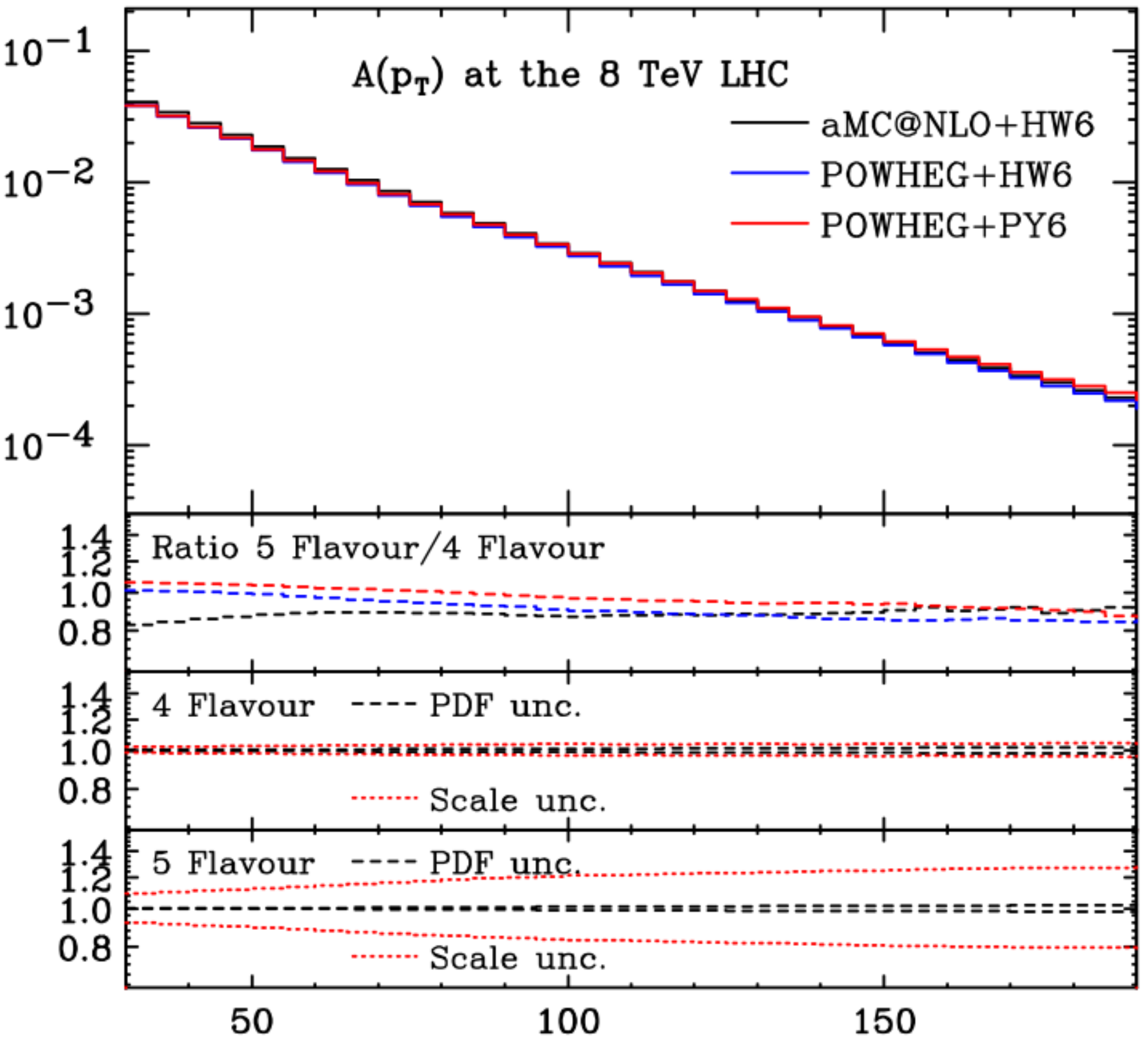} 
\end{minipage}
\begin{minipage}{0.48 \linewidth}
\vspace{24 mm}
\caption{
Acceptance as a function of the second-hardest $b$-jet at the $8\,$TeV LHC in the 4F scheme. Main plot: results for MC@NLO and POWHEG interfaced
to Herwig and Pythia. Lower insets: ratio of the results in the 5F and 4F scheme and scale and PDF dependence of the 4F and 5F prediction computed with 
MC@NLO. See \cite{Frederix:2012dh} for details.}
\label{fig:4F_VS_5F}
\end{minipage}
\end{center}
\end{figure}

Recently a comparison of the two NLO MCPS frameworks has also been performed in the four-flavour scheme \cite{Frederix:2012dh}.
Good agreement was again found between POWHEG and MC@NLO. As already observed for the fixed-order NLO
cross section \cite{Campbell:2009ss,Campbell:2009gj}, a comparison of results in the 4F and 5F schemes at the parton-shower level shows that the two descriptions are consistent at NLO, 
though the former gives more accurate predictions for observables which are related to the spectator $b$-jet. This can be seen for
example in Figure \ref{fig:4F_VS_5F}, where the acceptance $A(p_T)=1/\sigma \int^\infty_{p_T} d p_T^{(j_b,2)} \partial \sigma/\partial p_T^{(j_b,2)}$ 
 is shown as a function of the transverse momentum of the second-hardest $b$-jet. From the insets is clear that the 4F scheme result
has a much smaller theoretical uncertainty, though 4F and 5F predictions are consistent within their respective errors.        
 
\section{Conclusions}

In view of its phenomenological relevance, in the last few years a lot of effort has been put into providing an accurate theoretical description 
of single-top production at hadron colliders. State-of-the-art results are represented by fixed-order NLO predictions matched to Monte Carlo
parton showers, which provide an exact NLO description of the first emission and resum all-order radiation in the collinear region. In this 
framework the production and decay of the top quark is treated in the on-shell approximation. Studies of off-shell and non-factorizable effects
have been performed for $t$-channel and $s$-channel production, and they have shown that these effects, though most of the time small,  can be 
enhanced close to kinematical thresholds, and should be taken into account for a correct description of the shape of distributions close 
to kinematical edges. Resummation of Sudakov logarithms related to soft-gluon emission has also been studied by two different groups.
In both cases, small resummation effects were found for $t$-channel production, but somewhat contrasting results were presented for 
$s$-channel production, which would motivate a further investigation of these effects.           


\end{document}